\documentclass[aps,prb,twocolumn,showpacs,floatfix]{revtex4}

\usepackage{epsfig}

\begin{document}

\title{Ginzburg-Landau Vortex Lattice in
   Superconductor Films of Finite Thickness}

\author{E.~H.~Brandt}
\affiliation{Max-Planck-Institut f\"ur Metallforschung,
   D-70506 Stuttgart, Germany}
\date{\today}

\begin{abstract}
The Ginzburg-Landau equations are solved for ideally periodic
vortex lattices in superconducting films of arbitrary thickness
in a perpendicular magnetic field. The order parameter, current
density, magnetic moment, and the 3-dimensional magnetic field
inside and outside the film are obtained in the entire ranges
of the applied magnetic field, Ginzburg Landau parameter $\kappa$,
and film thickness. The superconducting order parameter varies
very little near the surface ($\approx 1\%$) and the energy of
the film surface is small. The shear modulus $c_{66}$
of the triangular vortex lattice in thin films coincides with
the bulk $c_{66}$ taken at large $\kappa$. In thin type-I
superconductor films with $\kappa < 1/\sqrt2$, $c_{66}$ can be
positive at low fields and negative at high fields.
The magnetization of thin films at small applied fields is
enhanced beyond its bulk value $-H_{c1}$ due to the
energy of the magnetic stray field.

\end{abstract}


\maketitle

\section{Introduction}

   Since Abrikosov's \cite{1} prediction of the flux-line lattice
in Type-II superconductors from Ginzburg-Landau (GL) theory, most
theoretical work on this vortex lattice in bulk and thin film
superconductors considered the situation when the applied magnetic
field and the average induction $\bar B$ are close to the upper
critical field $B_{c2} = \mu_0 H_{c2}$, since analytical solutions
may be obtained for this particular case. In the opposite limit
of low induction $\bar B \ll B_{c2}$, the properties of an isolated
vortex and the interaction between vortices are obtained to good
approximation from the London theory when the GL parameter $\kappa$
is not too small, $2\kappa^2 \gg 1$. \cite{1,2,3}
The problem of an isolated vortex in thin films was solved
analytically within London theory by Pearl \cite{4}; the
interaction energy of such Pearl vortices (or pancake vortices
\cite{5}) is easily calculated by noting that within London theory
the currents and magnetic fields of the vortices superimpose
linearly and that the force on a vortex equals the
thickness-integrated super-current density at the vortex core
times the quantum of flux $\Phi_0$. In thin films with thickness
$d$ smaller than the London magnetic penetration depth $\lambda$
the range of the vortex--vortex interaction is increased to the
effective penetration depth $\Lambda = \lambda^2/d$ since the
interaction now occurs mainly via the magnetic stray field
outside the film. \cite{4,5}
Vortices in superconducting films of finite thickness
($d < \lambda$ and $d \ge \lambda$) and in the superconducting
half space ($d \gg \lambda$) were calculated from GL
theory \cite{6} and London theory. \cite{7,8,9}

   At larger reduced induction $b = \bar B/B_{c2} > 0.05$ when
the London theory does not apply, the properties of the GL vortex
lattice have to  be computed numerically. A very efficient
method \cite{10} uses Fourier series as trial functions for the
GL function $|\psi(x,y)|^2$ and magnetic field $B(x,y)$ and
minimizes the GL free energy with respect to a finite number of
Fourier coefficients. This numerical method was recently
improved \cite{11,12} by solving the GL equations iteratively
with high precision.

  The present paper extends this two-dimensional (2D) method
to the 3D problem of a film of arbitrary thickness containing
a periodic lattice of GL vortices oriented perpendicular to the
film plane. Due to the Fourier ansatz, the magnetic
stray field energy is easily accounted for in this method.
Moreover, it turns out that the extension from 2D to a 3D
problem only slightly increases the required computation time
and computer memory, so that high precision can be achieved
easily on a Personal Computer. Like in Refs.~\onlinecite{11,12},
we consider here vortex lattices with arbitrary shape of
the unit cell containing one vortex, i.e., our method
computes triangular, rectangular, square lattices, etc.,
and yields also the shear moduli \cite{13} of the equilibrium
lattices. The approximate shear modulus $c_{66}$ of the
triangular vortex lattice in thin films was computed from GL
theory for $b \ll 1$ and $1-b \ll 1$ in Ref.~\onlinecite{14}.
For early work on films with perpendicular vortex lattice
see Refs.~\onlinecite{2,3,15,16,17,18,19}.

Though we consider here isotropic superconductors, the
corresponding results for anisotropic superconductors with
principal symmetry axes along $x,y,z$ may be obtained from this
isotropic method by scaling the coordinates and introducing
an effective GL parameter $\tilde \kappa$. \cite{20,21,22}
The magnetic field of a vortex inside a uniaxially
anisotropic superconductor with surface parallel to the
$a,c$ symmetry plane and perpendicular to the vortex line
was calculated from anisotropic London theory \cite{13}
and compared with experiments in Ref.~\onlinecite{23}.

  The main effect of the flat surface of a superconductor
film or half space is the widening of the magnetic field lines
of the vortices as they approach the surface. This widening
minimizes the sum of the bulk free energy plus the energy
of the magnetic stray field outside the superconductor.
The resulting magnetic field lines cross the superconductor
surface smoothly, see Fig.~1 for the vortex lattice and
Figs.~1,2 of Ref.~\onlinecite{9} for isolated vortices.
One can see that for the {\it vortex lattice} the
field lines at the boundary of the Wigner-Seitz cell are
exactly parallel to $z$, inside and outside the
superconductor, and at some distance outside from the
surface ($\approx$ half the vortex spacing) the magnetic field
becomes uniform and thus the field lines are parallel
and equidistant. For the {\it isolated vortex}, the
field lines away from the surface approach radial lines
as if they would originate from a point source, a  magnetic
monopole with strength $2\Phi_0$ situated on the vortex core
at a distance $1.27 \lambda$ below the surface. \cite{9}

  In Ref.~\onlinecite{6} the widening of the field lines
inside the superconductor was missed, but some modification
of the superconductor order parameter near the surface was
calculated from GL theory.  Below we obtain that the correct
modification of $|\psi|^2$ is very small: the vortex core,
visualized as contour lines of $|\psi(x,y,z)|^2$, widens near
the surface by only a few percent.

  The outline of this paper is as follows. In Sct.~2 the
solution method is outlined. Section 3 presents a selection
of results for thin and thick films:
Magnetic field lines, profiles of the order parameter and
magnetic field, the variances of the periodic order
parameter and magnetic field inside and outside the film,
surface energy and stray-field energy, and shear modulus of
the triangular vortex lattice in the film.
A summary is given in  Sct.~4.

 \begin{figure}  
\epsfxsize= .98\hsize  \vskip 1.0\baselineskip \centerline{
\epsffile{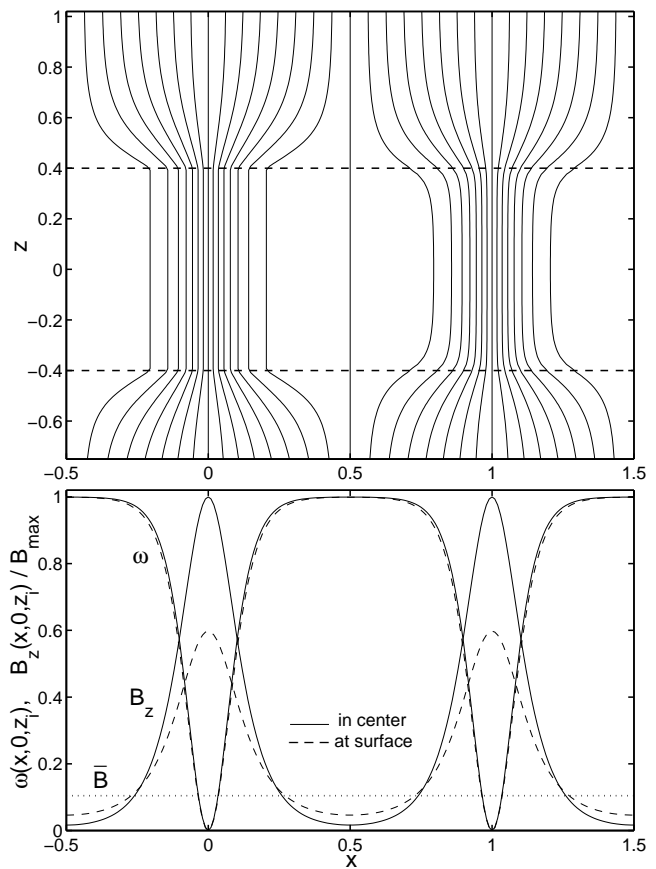}}  \vspace{.1cm}
\caption{\label{fig1} Magnetic field lines (top) and profiles of
order parameter $|\psi|^2 =\omega(x,0,z_i)$ and magnetic field
$B_z(x,0,z_i)$ (bottom) for a superconductor film calculated from
Ginzburg-Landau theory for the triangular vortex lattice.
Shown is the example
$b=\bar B/B_{c2} = 0.04$, $\kappa = 1.4$, triangular lattice
with vortex spacing (unit length) $x_1 =3^{-1/4}(2\Phi_0
  / \bar B)^{1/2} =5x_1(B_{c2}) \approx 10 \lambda$,
film thickness $d = 0.8 x_1 \approx 8 \lambda$.
Top: The left half shows the field lines that would apply if
the field in the film would not change near the surfaces
$z=\pm d/2$ marked by dashed lines. The right half shows the
correct solution. The density of the depicted field lines is
proportional to the local induction $|{\bf B}|$, see
Appendix A and Fig.~2.
Bottom: The solid lines show $\omega$ and $B$ in the center
of the film ($z=0$) and the dashed lines at the film surfaces.
The average induction $\bar B$ is marked as a dotted line.
 } \end{figure}   

 \section{Solution Method}

   The properties of the vortex lattice within GL theory are
calculated
by minimizing the GL free energy of the superconductor with
respect to the complex GL function $\psi({\bf r})$ and to the
vector potential ${\bf A}({\bf r})$ of the local magnetic
induction ${\bf B}({\bf r}) = \nabla \times {\bf A}$.
In the usual reduced units \cite{1,2}  (length
$\lambda$, magnetic induction $\sqrt2 \mu_0 H_c$, energy density
$\mu_0 H_c^2$, where $H_c = H_{c2}/(\sqrt2 \kappa)$ is the
thermodynamic critical field) the spatially averaged
free energy density $F$ of the GL theory
referred to the Meissner state ($\psi = 1$, ${\bf B}=0$)
within the superconductor reads
  \begin{equation}  
  F = \left\langle {(1-|\psi|^2)^2 \over 2} + \left| \left(
  {\nabla \over i\kappa} - {\bf A} \right) \psi \right|^2
  + {\bf B}^2 \right\rangle .
  \end{equation}
Here $\langle \dots \rangle = (1/V) \int_V d^3r \dots $ means
spatial averaging over the superconductor with volume $V$.
Introducing the supervelocity
${\bf Q}({\bf r}) = {\bf A} -\nabla\varphi/\kappa$ and the
magnitude $f({\bf r}) = |\psi|$ of
$\psi({\bf r}) = f({\bf r}) \exp[ i\varphi({\bf r}) ]$ one may
write $F$ as a functional of the real and gauge-invariant
functions $f$ and ${\bf Q}$,
  \begin{equation}  
  F = \left\langle {(1-f^2)^2 \over 2} + {(\nabla f)^2 \over
  \kappa^2 } +f^2 Q^2 +(\nabla\! \times\! {\bf Q})^2 \right\rangle.
  \end{equation}
In the presence of vortices ${\bf Q}({\bf r})$ has to be chosen
such that $\nabla\! \times\! {\bf Q}$ has the appropriate
singularities along the vortex cores, where $f$ vanishes.
By minimizing this $F$ with respect to $\psi$, ${\bf A}$ or
$f$, ${\bf Q}$, one obtains the GL equations together with the
appropriate boundary conditions.
For the superconducting film considered here, one has to add the
energy of the magnetic stray field
outside the film,  which makes  ${\bf B}$ continuous at the
film surface, see below.

 The 3D solution of the GL equations for an infinitely large,
thick or thin film with periodic lattice of vortices perpendicular
to the film plane, can be obtained numerically by a modification
of the 2D method developed in Refs. \onlinecite{11,12}. This is
possible since in any plane $z$ = const parallel to the film the
solutions for the ideal vortex lattice are still periodic.
Actually this periodicity applies even to tilted and arbitrarily
curved vortex lines, and to anisotropic superconductors, which
may be computed by a similar method. These more complex
problems will be considered in future work.

  For the present problem of straight vortices along $z$ one
may choose a general ansatz for the magnitude of the
GL function $f(x,y,z) = |\psi(x,y,z)|$ in form of the
following 3D Fourier series for the smooth function $f^2$:
  \begin{eqnarray}  
  \omega({\bf r}) =
  f^2 \! = \sum_{\bf K} a_{\bf K} (1 - \cos
  {\bf K}_\perp {\bf r}_\perp ) \cos K_z z \,.
  \end{eqnarray}
Here ${\bf r} = (x,y,z)$, ${\bf r}_\perp = (x,y)$,
${\bf K} = (K_x,K_y,K_z)$, ${\bf K}_\perp = (K_x,K_y)$.
In all sums here and below the term ${\bf K}_\perp =0$ is
excluded. For vortex positions
${\bf R} = {\bf R}_{mn} = (m x_1 +nx_2, \, ny_2)$
the reciprocal lattice vectors are
${\bf K_\perp} = {\bf K}_{mn} = (2\pi/S)(my_2, \,nx_1 +mx_2)$
with $S= x_1 y_2 = \Phi_0/\bar B$ the unit cell area and
$m = 0, \pm 1, \pm 2, \,\dots$ , $n = 0, \pm 1, \pm 2,
\,\dots$~. The $z$-component of ${\bf K}$ is chosen
as $K_z=(2\pi/d)l$ with $l = 0$, 1, 2, $\dots$ and $d$
the film thickness. This ansatz guarantees
that $f({\bf R},z) = 0$ at the (straight) vortex cores
and that at the two surfaces of the film $z = \pm d/2$ one
has $\partial (f^2) / \partial z = 0$, as it follows
from the variation of the GL free energy functional (2).
If only the term $K_z=0$ is kept, the ansatz (3) reduces
to that for the 2D vortex lattice in Ref. \onlinecite{1}.
Formally, the 3D Fourier series (3) may also be expressed
as a 2D Fourier series with $z$ dependent coefficients
$a_{{\bf K}_\perp}\!(z) =\sum_{K_z} a_{\bf K} \cos K_z z$.

   For the supervelocity ${\bf Q}$ and magnetic induction
${\bf B} =\nabla\! \times\! {\bf Q}$ inside the film we
chose the general ansatz
  \begin{eqnarray}  
   {\bf Q}({\bf r}) &=&{\bf Q}_A({\bf r}_\perp) +
                     {\bf q}({\bf r}) , \nonumber \\
   {\bf B}({\bf r}) &=& \bar B {\bf \hat z} + {\bf b}({\bf r}),
   ~~ \langle {\bf b}({\bf r}) \rangle = 0  ,
                                        \nonumber \\
   {\bf b}({\bf r}) &=& \nabla\! \times\! {\bf q}({\bf r}) \,.
  \end{eqnarray}
Here ${\bf Q}_A(x,y)$ is the supervelocity of the
Abrikosov $B_{c2}$ solution, which satisfies
  \begin{eqnarray}    
  \nabla\! \times\! {\bf Q}_A =\Big[ \bar B -\Phi_0 \sum_{\bf R}
  \delta_2({\bf r}_\perp \!-{\bf R}) \Big] {\bf\hat z}\,,
  \end{eqnarray}
where $\delta_2({\bf r}_\perp) = \delta(x)\delta(y)$ is
the 2D delta function and $\Phi_0$ the quantum of flux,
$\Phi_0 = 2\pi / \kappa$ in reduced units.
Formula (5) shows that ${\bf Q}_A$ is the
velocity field of a lattice of ideal vortex lines but with
zero average rotation. Near each vortex center one has
${\bf Q}_A({\bf r}_\perp) \approx {\bf\hat z \times r'}_\perp /
(2\kappa r'^2_\perp)$ and $f({\bf r})^2 \propto r'^2_\perp$
with ${\bf r'}_\perp= {\bf r}_\perp -{\bf R}$.
${\bf Q}_A({\bf r}_\perp)$ may be expressed as a slowly
converging Fourier series by integrating (5) using
${\rm div} {\bf Q} = {\rm div} {\bf Q}_A = 0$. It is, however,
more convenient to take ${\bf Q}_A$ from the exact relation
  \begin{eqnarray}    
  {\bf Q}_A({\bf r_\perp}) ={\nabla \omega_A \!\times {\bf\hat z}
  \over 2\, \kappa\, \omega_A } \,,
  \end{eqnarray}
where $\omega_A(x,y) = f(x,y)^2$ is the Abrikosov $B_{c2}$
solution given by a rapidly converging series of type (3) with
$z$-independent coefficients
  \begin{eqnarray}    
  a_{{\bf K}_\perp}^A = -(-1)^{m+mn+n} \exp[ -K_{mn}^2 S/(8\pi)]
  \end{eqnarray}
for general vortex-lattice symmetry, and
$a_{{\bf K}_\perp}^A = -(-1)^\nu \exp(-\pi \nu /\sqrt 3)$
($\nu = m^2 +mn +n^2$) for the triangular lattice. The $\omega_A$
from (7) is normalized to $\langle \omega_A(x,y) \rangle =1$;
this yields the strange relation
$\sum'_{{\bf K}_\perp} a_{{\bf K}_\perp}^A = 1$
for any lattice symmetry.
Another strange property of the Abrikosov solution (7) is that
$(\nabla\omega_A /\omega_A)^2 - \nabla^2 \omega_A /\omega_A =
 4\pi /S = {\rm const}$, although both terms diverge at the
vortex positions; this relation follows from (5) and (6)
using $\bar B = \Phi_0/S = 2\pi/(\kappa S)$. The useful
formula (6) may be proven via the complex
$B_{c2}$ solution $\psi_A(x,y)$; it means that near $B_{c2}$
the second and third terms in the $F$, Eq.~(2), are equal.

  The general ansatz for ${\bf q}({\bf r}) = (q_x, q_y, q_z)$
is a Fourier series for all three components, satisfying
$\nabla \times {\bf q} = {\bf b}$. For simplicity
here I shall assume $q_z =0$, which means planar supercurrents.
In the considered case of vortices perpendicular to
the film plane this is an excellent approximation, which is
exact in the limit of small induction and probably also
at large inductions $\bar B \approx B_{c2}$, and it is exact
for thin films.  I further assume
$\nabla\cdot {\bf Q} = 0$, which is exact in several special
cases (e.g. for $\bar B \ll B_{c2}$ and $\bar B \approx B_{c2}$)
and is possibly exact even in the general case, though I did not
find a proof for this. Note also that within the circular cell
approximation \cite{1,2} both assumptions are satisifed.
With these two assumptions ${\bf q}$ is fully determined by
the $z$-component of ${\bf b} = ({\bf b}_\perp, b_z)$:
  \begin{eqnarray}  
  b_z({\bf r}) &=& \sum_{\bf K} b_{\bf K} \cos
  {\bf K}_\perp {\bf r}_\perp  \cos K_z z \,,
           \nonumber \\ \nonumber
  {\bf b}_\perp({\bf r}) &=& \sum_{\bf K} b_{\bf K}
      {{\bf K}_\perp K_z \over K_\perp^2}
  \sin{\bf K}_\perp {\bf r}_\perp  \sin K_z z \,
            \\
  {\bf q}({\bf r}) &=& \sum_{\bf ~K} b_{\bf K}
  { {\bf \hat z \times K}_\perp \over K^2_\perp }
  \sin{\bf K}_\perp {\bf r}_\perp \cos K_z z \,,
  \end{eqnarray}
with $K_\perp = |{\bf K}_\perp |$.
Formally, these 3D Fourier series (8) may also be expressed
as 2D Fourier series with $z$ dependent coefficients
$b_{{\bf K}_\perp}\!(z) = \sum_{K_z} b_{\bf K} \cos K_z z$
and their derivatives $b_{{\bf K}_\perp}'\!(z)$. The solution
is thus completely determined by the two infinite
sets of scalar Fourier coefficients $a_{\bf K}$ and
$b_{\bf K}$, which are obtained by minimizing the total free
energy with respect to these coefficients for given
parameters $\kappa$ and $\bar B$ and film thickness $d$.
For the computation I shall use a large but finite number of
$a_{\bf K}$ and $b_{\bf K}$ in the sense of a Ritz
variational method.

   The total free energy $F_{\rm tot}$ per unit volume of the
infinite film is the sum of the GL free energy, Eq.~(2), and
the stray-field energy $F_{\rm stray}$. In reduced units
and  referred to the state where $\psi=0$ and
${\bf B(r)} = \bar B {\bf \hat z} = \mu_0 H_a {\bf \hat z}$
one has with $g = (\nabla f)^2 /\kappa^2 =
  (\nabla \omega)^2/(4\kappa^2 \omega)$:
  \begin{eqnarray}  
  &&F_{\rm tot}   = \langle -\omega
  + {\textstyle \frac{1}{2}} \omega^2
  +g +\omega Q^2 +b^2 \rangle + {F_{\rm stray} \over d} ,~~~
                \nonumber \\
  &&F_{\rm stray} =  2\! \int_{d/2}^\infty
  \langle\, {\bf B(r)}^2 - \bar B^2 \,\rangle_{x,y} \, dz .
  \end{eqnarray}
The factor of 2 comes from the two half spaces above and
below the film, which contribute equally to $F_{\rm stray}$.
The stray field ${\bf B}(x,y,z>d/2)$ with constant planar
average
$\langle {\bf B}(x,y,z) \rangle_{x,y} = \bar B{\bf \hat z}$ is
determined by the Laplace equation $\nabla^2 {\bf B} =0$
(since $\nabla\cdot {\bf B} =0$ and $\nabla \times {\bf B}=0$
in vacuum) and by its perpendicular component at the film
surface $z= d/2$, since $B_z$ has to be continuous across
this surface. This yields with (8) the stray field:
  \begin{eqnarray}  
  B_z(x,y,z \ge d/2 )  = \nonumber \\
   \sum_{{\bf ~K}_\perp}   b_{{\bf K}_\perp}^s \!\!
   \cos( {\bf K}_\perp {\bf r}_\perp)
   \exp[ -K_\perp (z-d/2 ) ]  + \bar B\,,
       \nonumber \\ \nonumber
 {\bf B}_\perp (x,y,z \ge d/2 )  = \nonumber \\
   \sum_{{\bf ~K}_\perp}   b_{{\bf K}_\perp}^s \! {{\bf K}_\perp
   \over K_\perp } \sin( {\bf K}_\perp {\bf r}_\perp )
   \exp[ -K_\perp (z-d/2 ) ] \,,
       \nonumber \\
  b_{{\bf K}_\perp}^s \! =  b_{{\bf K}_\perp}\! (z=d/2) =
     \sum_l b_{\bf K} \cos(\pi l) \,.
  \end{eqnarray}
($l=0, 1, 2, \dots$). For spatial averaging we shall need the
orthonormality relations valid for ${\bf K}_\perp \ne 0$:
  \begin{eqnarray}  
   \langle\, \cos( {\bf K}_\perp {\bf r}_\perp)
   \cos( {\bf K'}_\perp {\bf r}_\perp) \,\rangle_{x,y}
   &=&    \nonumber \\
   \langle\, \sin( {\bf K}_\perp {\bf r}_\perp)
   \sin( {\bf K'}_\perp {\bf r}_\perp) \,\rangle_{x,y}
   &=& {1\over2}\,\delta_{{\bf K}_\perp {\bf K'}_{\!\!\perp}} ,
  \end{eqnarray}
  \begin{eqnarray}  
   \left\langle\, \cos{2\pi lz\over d} \cos{2\pi l'z \over d} \,
   \right\rangle_{\!z}=\delta_{l,l'} {1 +\delta_{l,0} \over 2}\,,
                               \nonumber \\
   \left\langle\, \sin{2\pi lz\over d} \sin{2\pi l'z \over d} \,
   \right\rangle_{\!z}=\delta_{l,l'} {1 -\delta_{l,0} \over 2}\,.
  \end{eqnarray}
Averaging the squared stray field over $x$ and $y$ and using
(11), (12) one obtains terms
$( b_{{\bf K}_\perp}^s\! )^2  \exp[ -2K_\perp (z-d/2 ) ]$, and
thus $F_{\rm stray}$ in Eq.~(9) becomes
  \begin{eqnarray}  
   F_{\rm stray} = \sum_{{\bf ~K}_\perp}
   { (\,b_{{\bf K}_\perp}^s \!)^2  \over ~K_\perp } \,.
  \end{eqnarray}
The Fourier coefficients $a_{\bf K}$ and $b_{\bf K}$ may
be computed by iterating appropriate GL equations as shown
in Ref.\ \onlinecite{11,12}. Minimizing $F$, Eq.~(2), with
respect to $f$ and ${\bf Q}$ yields the two GL equations
for bulk superconductors
  \begin{eqnarray}  
   \kappa^{-2} \nabla^2 f = -f +f^3 +fQ^2  \,,\\
   {\bf j} = \nabla \times {\bf B} = \nabla \times \nabla
   \times {\bf Q} = -f^2 {\bf Q} \,.
  \end{eqnarray}
The first GL equation (14) applies also to our film;
the second GL equation (15), too, but when it is written as an
equation for the $b_{\bf K}$ it has to be supplemented by a
stray-field term $\sim \delta F_{\rm stray} / \delta {\bf Q}$
on its r.h.s., which originates from the boundary condition for
${\bf B}$. A possible iteration equation for the $a_{\bf K}$
is obtained from (14) using the relation
$2 f \nabla^2 f = \nabla^2 \omega -(\nabla \omega)^2/(2\omega)$
to give
  \begin{eqnarray}  
  \nabla^2 \omega =2\kappa^2 \,
  (-\omega +\omega^2 +\omega Q^2 +g \,)
  \end{eqnarray}
with $g = (\nabla \omega)^2/(4\kappa^2 \omega)$ as above.
Note that $\nabla$ here means the 3D Nabla operator, while the
similar Eq.~(9) of Ref.\ \onlinecite{11} is 2D. To obtain better
convergence of the iteration I subtract a term $2\kappa^2 \omega$
on both sides of (16), such that $K^2$ is replaced by
$K^2 + 2\kappa^2$;  this choice yields fastest convergence.
Using the ansatz (3) and the orthonormalities (11), (12)
we then obtain an iteration equation for the $a_{\bf K}$:
  \begin{eqnarray}  
  a_{\bf K} := {\langle (\omega^2\! -\!2\omega + \omega Q^2\!+\!g)
   \, \cos{\bf K}_\perp {\bf r}_\perp \cos K_z z \rangle \over
   {1\over 4}\,( \delta_{K_z,0} +1)\, (K^2/2\kappa^2 +1) } \,,~~~
  \end{eqnarray}
where $\langle \dots \rangle$ averages over $x,y,z$.
In particular, if $\omega$ and ${\bf Q}$ do not depend on $z$,
Eq.~(17) reduces to Eq.~(11) of Ref. \onlinecite{11} and
yields $a_{\bf K} = 0$ for all $K_z \ne 0$.
%
Other forms of iteration equations for the $a_{\bf K}$ are
possible, e.g. one containing in the denominator
$K_\perp^2$ instead of $K^2$, but one should choose that which
yields fastest convergence of the iteration.
In general, the iteration of any equation for some
parameter $a$ given in the original form $a:= F(a)$ may be made
more stable or faster converging by rewriting it in the form
$a:= (1-c)a +c F(a)$ with some constant $c \le 1$ (or even
$c>1$ in some cases).

   The convergence is accelerated by alternating the
iteration step (17) with an iteration step that changes
only the amplitude of $\omega$ but not its shape. Namely, from
$\partial F_{\rm tot} / \partial \langle \omega \rangle =0$
we obtain \\[-6mm]
  \begin{eqnarray}  
   a_{\bf K} := a_{\bf K} {\langle \omega - g - \omega Q^2
   \rangle \over \langle \omega^2 \rangle } \,.
  \end{eqnarray}

   Similarly, an iteration equation for the $b_{\bf K}$
is obtained from the equation
$\partial F_{\rm tot} / \partial b_{\bf K} =0$ by reordering
the terms appropriately. From Eq. (8)-(13) one has
  \begin{eqnarray}  
  { \partial \langle b^2 \rangle \over \partial b_{\bf K} }
  &=& b_{\bf K} { (1\! +\! \delta_{l,0})K_\perp^2 +
      (1\! -\! \delta_{l,0}) K_z^2 \over 2 K_\perp^2 } ~~~\\
  { \partial \langle \omega Q^2 \rangle \over \partial b_{\bf K}}
  &=& { 2 P_{\bf K} \over K_\perp^2 }
        \\    {1\over d}
  { \partial F_{\rm stray} \over \partial b_{\bf K}}
  &=& b_{{\bf K}_\perp}^s {2 \cos(\pi l) \over ~d\,K_\perp }
  \end{eqnarray}
with $ b_{{\bf K}_\perp}^s $ from Eq.~(10) and   
   \begin{eqnarray}  
  P_{\bf K} = \langle\, \omega (Q_y K_x \! -Q_x K_y)
   \sin {\bf K}_\perp {\bf r}_\perp \cos K_z z \,\rangle\,.~
 \end{eqnarray}  \\[-2mm]
Equating the sum of the terms (19)--(21) to zero and adding
and subtracting an appropriate term
$c \langle \omega \rangle b_{\bf K}$ that improves the
convergence (with some constant $c \approx 1$ or larger), one
obtains an iteration equation for the $b_{\bf K}$ :
  \begin{eqnarray}  
  b_{\bf K} :={-2P_{\bf K}+ c\langle \omega \rangle\, b_{\bf K}
  - {2 \over d} K_\perp  b_{{\bf K}_\perp}^s \! \cos(\pi l)
  \over \delta_{l,0} K_\perp^2 +{1\over 2}\,(1 -\delta_{l,0})
  K^2 + c\langle \omega \rangle } \,.
 \end{eqnarray}

The solutions $\omega({\bf r})$, ${\bf B(r)}$, and ${\bf Q(r)}$
are obtained iteratively by first finding the 2D solution as in
Ref.~\onlinecite{11,12}, keeping only the terms with $K_z=0$ and
starting, e.g., with
$a_{\bf K} = (1- \bar B/B_{c2})\, a_{{\bf K}_\perp}^A $
and $b_{\bf K} = 0$ and then iterating the three equations
(17), (18), (23) by turns a few times; after this, the 3D solution
is obtained by continuing this iteration with the terms for all
$K_z$ included until the coefficients $a_{\bf K}$ and $b_{\bf K}$
do not change any more. With the empirical choice
$c \approx 3 + (0.4 +60 b^2)\kappa^2 x_1/ d$ this iteration is
stable for all $b$, $\kappa$, and $d$ and the free
energy decreases smoothly until it becomes stationary (with
accuracy $10^{-14}$) after $25 \dots 50$ iteration steps.


 \begin{figure}  
\epsfxsize= .98\hsize  \vskip 1.0\baselineskip \centerline{
\epsffile{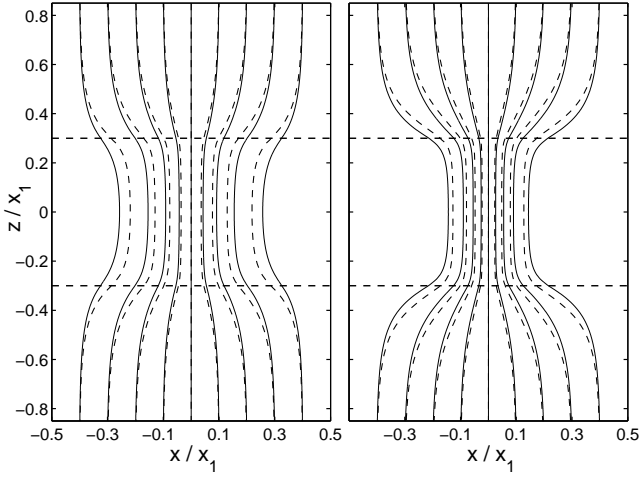}}  \vspace{.1cm}
\caption{\label{fig2} Comparison of the magnetic field lines
plotted either as stream lines (solid lines) that flow along
the exact direction of the magnetic field but do not show the
correct 2D flux density, or as contour lines (dashed lines,
see Appendix A) that show the correct flux density but have
only approximately the orientation the magnetic field. Shown
are the examples $b=0.04$, triangular lattice, with (right plot)
$\kappa = 2$, $d = 0.6 x_1 \approx 4 \lambda$ and (left plot)
$\kappa = 1$, $d = 0.6 x_1 \approx 8 \lambda$. The horizontal
dashed lines indicate the film surfaces $z=\pm d/2$.
 } \end{figure}   

 \begin{figure}  
\epsfxsize= .98\hsize  \vskip 1.0\baselineskip \centerline{
\epsffile{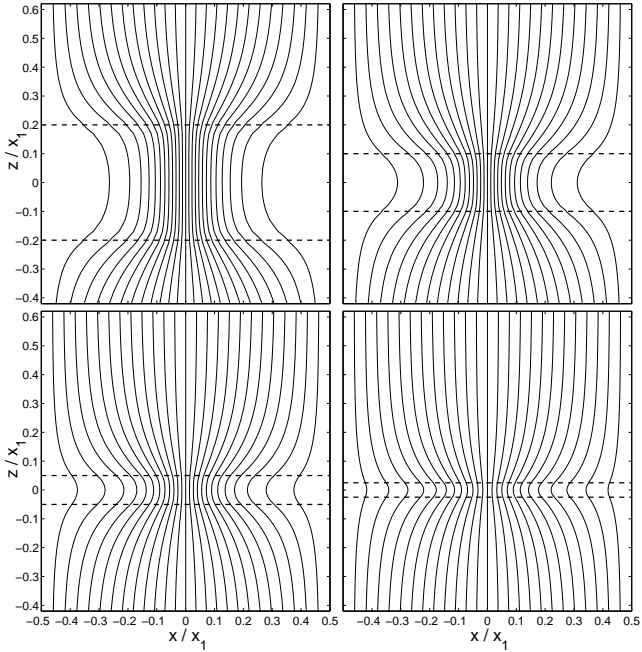}}  \vspace{.1cm}
\caption{\label{fig3} The magnetic field lines of the vortex
lattice in films of various thicknesses
$d/\lambda \approx 4$, 2, 1, 0.5, corresponding to
$d / x_1 = 0.4$, 0.2, 0.1, 0.05, for $b=0.04$ and $\kappa =1.4$.
Depicted is the field in the plane $y=0$ in one lattice cell.
The film surfaces are marked by two dashed lines.
The field lines of an isolated vortex in such
films are shown in Fig.~2 of Ref.~9.
 } \end{figure}   

 \begin{figure}  
\epsfxsize= .98\hsize  \vskip 1.0\baselineskip \centerline{
\epsffile{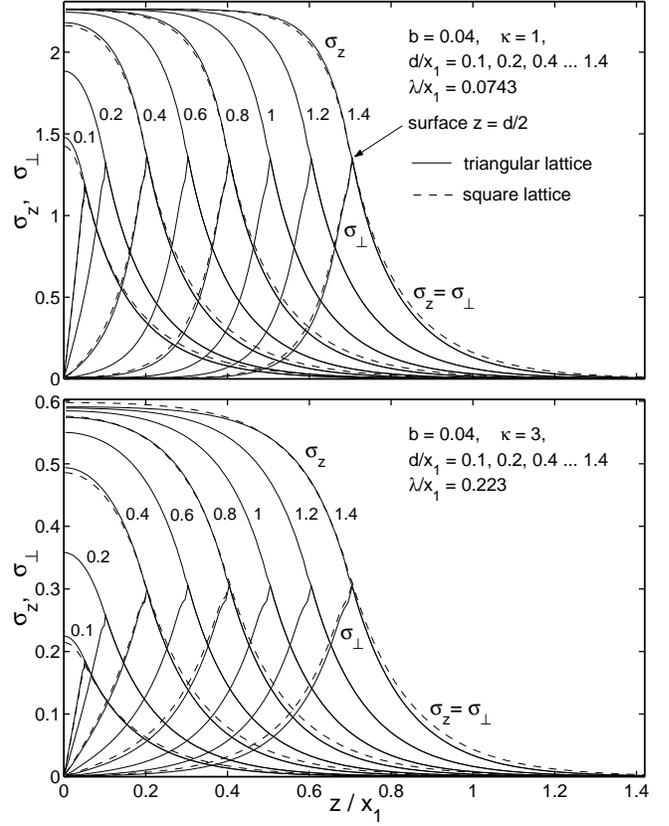}}  \vspace{.1cm}
\caption{\label{fig4} The variances of the longitudinal and
transverse components of the magnetic induction,
$\sigma_z(z)$ and $\sigma_\perp(z)$, defined by Eqs.~(24), (25),
plotted versus $z/x_1$ ($x_1$ = vortex spacing) at
reduced induction $b=0.04$ for films of various thicknesses
$d/x_1 = 0.1$, 0.2, 0.4, 0.6, 0.8, 1, 1.2, and 1.4.
Top: For $\kappa = 1$, yielding $\lambda/x_1 = 0.0743$.
Bottom: For $\kappa = 3$, thus  $\lambda/x_1 = 0.223$.
While $\sigma_z(z)$ decreases monotonically with increasing
$z$,  $\sigma_\perp(z)$ has a sharp peak at the film surface
$z = d/2$. Outside the film ($|z| \ge d/2$) one has
$\sigma_z(z) = \sigma_\perp(z)$.
 } \end{figure}   

 \begin{figure}  
\epsfxsize= .98\hsize  \vskip 1.0\baselineskip \centerline{
\epsffile{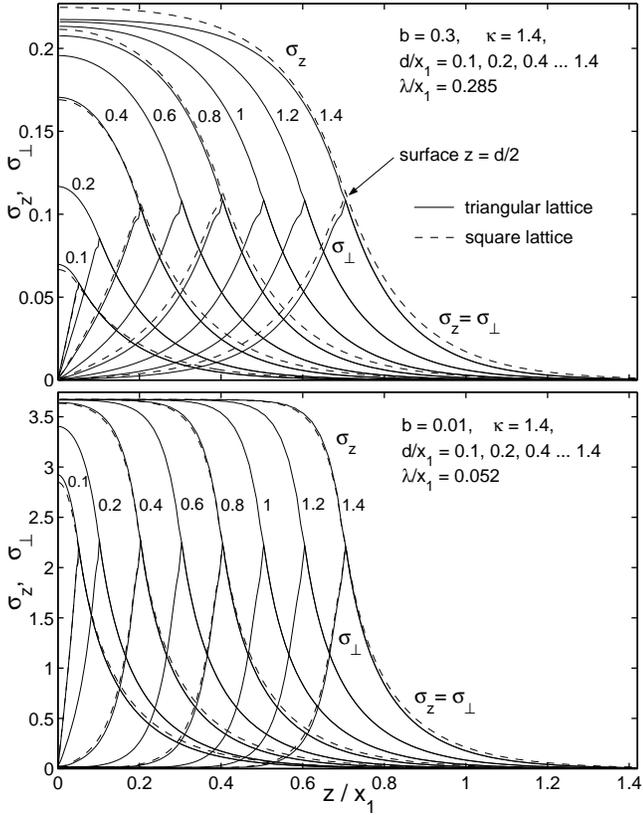}}  \vspace{.1cm}
\caption{\label{fig5} As Fig.~4 but for $\kappa = 1.4$ and
two $b$ values.
Top: Large $b=0.3$, yielding $\lambda/x_1 = 0.285$.
Bottom: Low $b=0.01$, thus   $\lambda/x_1 = 0.0520$.
 } \end{figure}   

 \begin{figure}  
\epsfxsize= .98\hsize  \vskip 1.0\baselineskip \centerline{
\epsffile{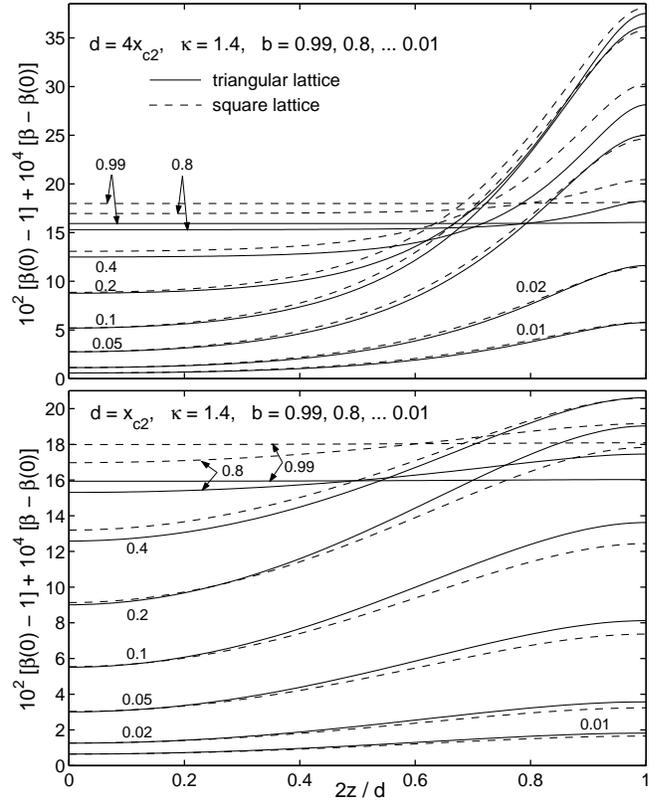}}  \vspace{.1cm}
\caption{\label{fig6} The spatial variance (Abrikosov
parameter) $\beta(z) = \langle\, \omega^2 \,\rangle_{x,y} /
            \langle\, \omega \,\rangle_{x,y}^2$,
of the order parameter $\omega(x,y,z) = |\psi(x,y,z)|^2$,
plotted as $10^4 [\beta(z) -\beta(0] ] +100 [\beta(0) -1]$
versus $z/(d/2)$ for several reduced inductions
$b = 0.99$, 0.8, 0.4, 0.2, 0.1, 0.05, 0.02, and 0.01
for $\kappa=1.4$ and for both triangular (solid lines)
and square (dashed lines) vortex lattices.
Top: Thick film with $d = 4 x_{c2}$.
Bottom: Thinner film with $d = x_{c2}$. Here
$x_{c2} = x_1 \sqrt b \approx 2.7 \xi$ is the vortex spacing
at $\bar B = B_{c2}$. The extremely small variation of
$\beta(z)$ is enlarged by plotting
$10^4 \times [\beta(z) - \beta(0)]$. Adding the constant
$10^2 [\beta(0)-1]$ allows to identify (at $z=0$) the bulk
Abrikosov values $\beta(0) \approx \beta_A = 1.1596$ (1.8034)
occurring for the triangular (square) lattice in
thick films when $b \to 1$.
 } \end{figure}   

 \begin{figure}  
\epsfxsize= .98\hsize  \vskip 1.0\baselineskip \centerline{
\epsffile{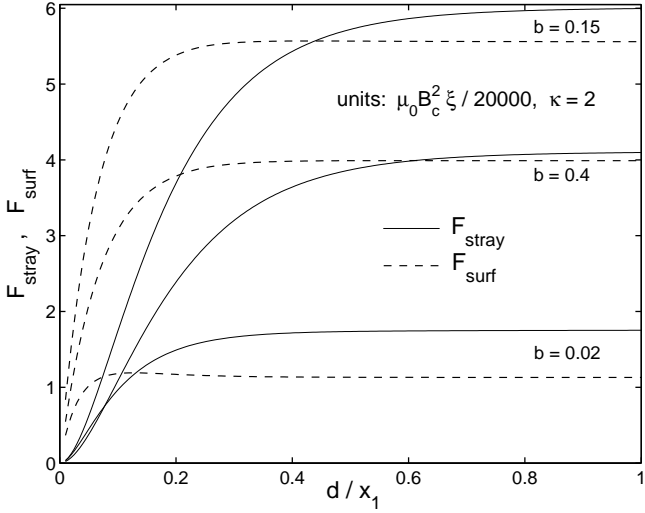}}  \vspace{.1cm}
\caption{\label{fig7}
  The energy $F_{\rm stray}$ of the magnetic stray field
(solid lines) and the surface energy $F_{\rm surf}$ of the
film (dashed lines) plotted versus the film thickness $d$ for
$\kappa =2$ and $b=0.02$, 0.15, and 0.4, see text. These
energies per unit area are plotted in units
$\mu_0 H_c^2 \xi / 20000$ to enlarge them to order of
unity and show their close coincidence at large $b$.
 } \end{figure}   

 \begin{figure}  
\epsfxsize= .98\hsize  \vskip 1.0\baselineskip \centerline{
\epsffile{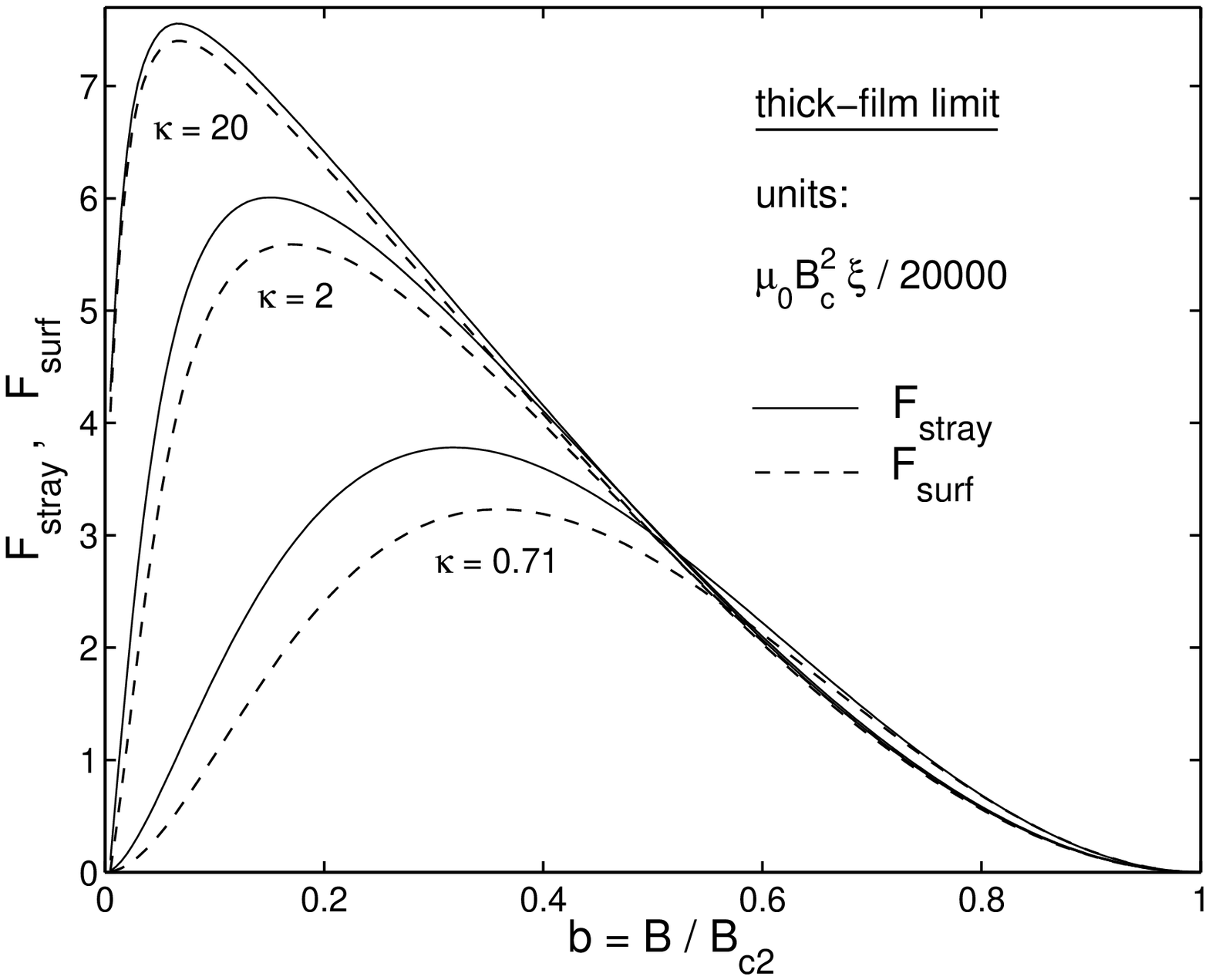}}  \vspace{.1cm}
\caption{\label{fig8}
The thick-film limits of the stray-field energy $F_{\rm stray}$
(solid lines) and the surface energy $F_{\rm surf}$ (dashed lines)
plotted versus the reduced induction $b$ for $\kappa=0.71$, 2,
and 20 in units $\mu_0 H_c^2 \xi / 2\cdot 10^4$.
 } \end{figure}   

\section{Some results}  

\subsection{Magnetic field and order parameter}  

  Figure 1 shows
  one example for the resulting  magnetic field lines and
some  cross sections of $\omega(x,y,z)$ and ${\bf B}(x,y,z)$
along $x$ in the plane $y=0$ at $z=0$ (center plane of the film)
and $z=d/2$ (film surface), for a film of finite thickness
$d = 0.8 x_1 \approx 8 \lambda$ at reduced induction
$b=\bar B / B_{c2} = 0.04 $ and GL parameter $\kappa = 1.4 $,
yielding for the triangular vortex lattice a vortex spacing
of $x_1(B) = 5x_1(B_{c2}) = 1.25 d \approx 10 \lambda$.
The left half of Fig.~1 shows the field lines that result
if the unchanged 2D bulk solutions for $B(x,y)$ and
$\omega(x,y)$ are assumed inside the film. The right half
shows the correct solution, exhibiting smooth field lines
across the film surface, and a very weak widening of the
vortex core near the surface.

  Figure 2 shows the magnetic field lines for a film with
thickness $d = 0.6 x_1$ at $b=0.04$ for $\kappa = 2$
($d \approx 4 \lambda$, left) and $\kappa = 1$
($d \approx 8 \lambda$, right), triangular lattice.
The solid lines are the stream lines of
${\bf B}(x, 0, z) = (B_x, 0, B_z)$; they have the correct
slope of ${\bf B}$ and start at equidistant points far
away from the film surface, where
${\bf B} = {\bf B}_a = (0,0,\bar B)=$ constant
(in infinitely extended films the average induction
${\bf \bar B}$ equals the applied induction ${\bf B}_a$
outside and inside the film),
but their 1D density is not proportional to the
2D flux density $B = |{\bf B}|$. The dashed lines in
Fig.~2 are field lines that have approximately the slope
of ${\bf B}(x, 0, z)$ and have a density proportional to
$B$, see Appendix A. This type of field lines is
depicted also in Figs.~1 and 3.

  In Fig.~3 the magnetic field lines are shown for films
of various thicknesses $d/x_1 = 0.4$, 0.2, 0.1, 0.05
for $b=0.04$ and $\kappa = 1.4$ as in Fig.~1, where
$d/x_1 = 0.8$. These thicknesses correspond to
$d/\lambda \approx 4$, 2, 1, 0.5 (and 0.25 in Fig.~1).
At low inductions $b \ll \kappa^{-2}$ and not too small
$\kappa > 5$, these field patterns may also be obtained by
linear superposition of the fields of isolated London
vortices given by Eqs.~(5)-(9) of Ref.~\onlinecite{9}, with
appropriately cut-off vortex core introduced to consider
the finite coherence length $\xi$. This superposition also
applies to nonperiodic vortex arrangements.

\subsection{Variance of the magnetic induction}  

  Figures 4 and 5 show the two relative  variances
$\sigma_z$ and $\sigma_\perp$ of the magnetic
induction defined here as
  \begin{eqnarray}   
  \sigma_z(z)  &=& \langle\, [ B_z(x,y,z) -\bar B ]^2
               \,\rangle_{x,y}^{1/2} / \bar B  \\
  \sigma_\perp(z)&=& \langle\,B_x(x,y,z)^2\,+B_y(x,y,z)^2
               \,\rangle_{x,y}^{1/2} / \bar B  \,.
  \end{eqnarray}
These measures of the relative variation of the periodic
induction depend on $z$: Deep inside thick films,
$\sigma_z(z)$ reduces to the variance of the 2D vortex
lattice in the bulk, $\sigma_{\rm bulk}$, computed, e.g.,
in Ref.~\onlinecite{12} as function of $b$ and $\kappa$,
and one has $\sigma_\perp = 0$ since
${\bf B} \| {\bf \hat z}$ for the considered case.
As one approaches the surface from inside the film,
$\sigma_z(z)$ decreases and $\sigma_\perp(z)$ increases
until they coincide at the surface $z=d/2$. Outside the
superconductor one has exactly
  \begin{eqnarray}   
  \sigma_z^2(z \ge d/2) = \sigma_\perp^2(z \ge d/2)
     ~~~~~~~  \nonumber \\
   = {1 \over 2 \bar B^2} \sum_{{\bf ~K}_\perp}
   (\,b_{{\bf K}_\perp}^s \!)^2 \exp[-2K_\perp(z-d/2)] \,.
  \end{eqnarray}
This follows from Eqs.~(10) for the magnetic
stray field. At large $|z| -d/2 \gg x_1/(2\pi)$, the
variance decreases exponentially with $z$,
$ \sigma_z = \sigma_\perp \propto \exp(-K_{10}|z|)$,
where $K_{10} \approx 2\pi /x_1$ is the shortest reciprocal
lattice vector of the vortex lattice defined below Eq.~(3).
Thus, $\sigma_z(z)$ decreases monotonically from its
bulk value $\sigma_{\rm bulk}$ inside a thick film to
zero far away from the film, reaching at the surface
approximately half its bulk value (for thick films).
In contrast to this, the transverse variance
$\sigma_\perp(z)$ increases from zero at $z=0$ and
reaches a sharp cusp-shaped maximum at the surface,
where it joins $\sigma_z(z)$ and then decreases again
to zero away from the film.
For large $\kappa \ge 2$ and $d/x_1 \ge 0.7$ these
curves are approximately symmetric,
  \begin{eqnarray}   
  \sigma_z(|z| < d/2) = \sigma_{\rm bulk} -\sigma_z(d-|z|)
         \nonumber \\
  \sigma_\perp(|z| < d/2) \approx \sigma_z(d-|z|) ~~~~~
         \nonumber \\
  \sigma_\perp(d/2) = \sigma_z(d/2) \approx
                      \sigma_{\rm bulk}/2 ~.~~
    \end{eqnarray}
This is so since for $\lambda \gg x_1/2\pi$ the outer
and inner magnetic stray fields are symmetric.

   Shown in Figs.~(4) and (5) are both variances for
various values of $b$ and $\kappa$ for 8 film thicknesses
$d/x_1$ = 0.1, 0.2, 0.4, 0.6, 0.8, 1, 1.2, and 1.4 for
the triangular vortex lattice (solid lines) and for
$d/x_1 = 0.1$, 0.4, 0.8, and 1.2 for the square vortex
lattice (dashed lines). The variances for these two
lattice symmetries are almost identical.

\subsection{Variance of the order parameter}  

   Figure 6 shows the variance of the order parameter
$\omega(x,y,z) = |\psi(x,y,z)|^2$ inside films with
periodic vortex lattice,
  \begin{eqnarray}   
  \beta(z) = {\langle\, \omega(x,y,z)^2 \,\rangle_{x,y}
        \over \langle\, \omega(x,y,z) \,\rangle_{x,y}^2}.
   \end{eqnarray}
At large reduced inductions $b = \bar B/B_{c2} \to 1$ in
the middle of thick films $\beta(z)$ coincides with the
Abrikosov parameter $\beta_A = 1.15960$ for the triangular
vortex lattice and  $\beta_A = 1.18034$ for the square
lattice. At low inductions $b \to 0$ one has $\beta\to 1$
since the order parameter $\omega$ is constant
except in the small vortex cores.
Figure 6 shows that $\beta(z)$, and thus the shape of
$\omega(x,y,z)$,
changes very little with $z$. For films of thickness
$d > 4 x_{c2}$, one has constant bulk $\beta$ in a central
region around $z=0$, and as $z$ approaches the surface of
the film, $\beta(z)$ increases by at most a factor of
1.0033 within a layer of thickness $\approx x_{c2}$. Here
$x_{c2} = x_1(B_{c2}) = x_1 \sqrt b$ is the vortex distance
at $\bar B=B_{c2}$. One has $x_{c2}^2 =(4\pi/\sqrt 3)\xi^2$
for the triangular and $x_{c2}^2 = 2\pi\xi^2$ for the
square vortex lattice. The maximum change occurs at
$b \approx 0.15$ and $d \ge 4x_{c2}$.
For thinner films and larger and smaller $b$, the
variation of $\beta(z)$ is even smaller.

  Thus, to a very
good approximation, one may assume that the order
parameter $\omega(x,y,z)$ of the vortex lattice inside
films in perpendicular magnetic field is independent of
$z$ and for not too thin films has the same form as the 2D
order parameter $\omega(x,y)$ of the bulk vortex lattice.
For very thin films with $d \ll \xi$ at $b \ll 1$ the vortex
cores are slightly wider than in the bulk. For example, at
$b=0.04$, $\kappa = 1.4$, the core width increases by about
25 \% when $d/\xi$ decreases from 0.5 to 0.005, but then
saturates and does not increase further in thinner films.
This is just the interval of $d$ in which the modulation
$1 - B_{\rm min}/B_{\rm max}$ of the periodic magnetic field
$B(x,y,0)$ decreases from $\approx 1$ to $\ll 1$ since the
effective penetration depth $2\lambda^2 /d$ becomes larger.

   Close to $B_{c2}$ the constancy along $z$ of the
GL function $\psi(x,y,z)$ applies to thicker and
thicker films. This numerical result is consistent
with the finding in Ref.~\onlinecite{24} of a
correlation length $l_z = \xi/(2\sqrt{1-b})$ that
diverges for $b \to 1$ and describes the extension
along the vortex lines of perturbations in $\psi(x,y,z)$
caused by small material inhomogeneities (pins).
Interestingly, a similar diverging length
$\xi/(2\sqrt{1-b})$ describes the long axis (along $z$)
of a cigar-shaped superconducting region (nucleus) that
nucleates at applied fields above $B_{c2}$ at a small
defect with transition temperature $T_c({\bf r})$
higher than the bulk $T_c$. \cite{19}
In superconducting films of thickness
$ d < \xi |b-1|^{-1/2}$, or at applied fields
satisfying $|B_a/B_{c2} -1| < \xi^2 / d^2$, small
inclusions or precipitates are thus expected to cause a
virtually 2D perturbation that has no $z$-dependence.

\subsection{Surface energy}  

   Next I consider the additional free energy caused by
the presence of the two surfaces of the film. This energy
per unit area of the film is composed of two terms:

   (a) $F_{\rm stray}$, the  magnetic energy of the
stray field outside the film, defined by Eq.~(9) and expressed
in Eq.~(13) in terms of the Fourier coefficients
$b_{{\bf K}_\perp}^s$ of the field component
$B_z(x,y, d/2)$ at the surface,

   (b) $F_{\rm surf}$, the actual surface energy
defined as the difference of the free energy of the
film per unit area, $F_{\rm 3D}d$, minus the 2D bulk
free energy density of the infinite vortex lattice,
$F_{\rm 2D}$, times $d$, thus
  \begin{equation}  
    F_{\rm surf} = (F_{\rm 3D} -F_{\rm 2D})d.
  \end{equation}

The total surface energy, originating from both surfaces
of the film, is the sum of these two terms,
$F_{\rm stray} +F_{\rm surf}$. Both terms tend to a
constant when the film thickness $d$ increases above the
vortex spacing $x_1$. These thick-film values of
$F_{\rm stray}$ and $F_{\rm surf}$ are of the same order,
and they are approximately equal for large $\kappa \gg 1$
and also at large reduced inductions $b \to 1$. This
is so since the order parameter $\omega(x,y,z)$ in the
film is nearly independent of $z$, and thus $F_{\rm surf}$
is virtually only of magnetic origin, i.e., it is the
energy of the magnetic field change caused inside the
film by the surface. When the magnetic penetration depth
is large,  $\lambda > x_1/2\pi \ll d$, this
 ``inner stray field'' is symmetric to the outer stray field.
This equality applies also near $b=1$, since inside the
superconductor the magnetic screening is reduced by the
reduction of the order parameter and thus the effective
penetration depth $\lambda' = \lambda / \langle \omega
\rangle ^{1/2} \approx \lambda / (1-b)^{1/2}$
increases. \cite{13}

  The dependence of $F_{\rm stray}$ and $F_{\rm surf}$
on the film thickness is shown in Fig.~7 for $\kappa = 2$
and $b=0.02$, 0.15, and 0.4. With increasing $d$ both
energies increase from zero and saturate to constant values
at about $d/x_1 \ge 0.7$ for all $\kappa$ and $b$.
Figure 8 shows these thick-film limits of $F_{\rm stray}$
and $F_{\rm surf}$ as functions of the reduced induction
$b$. Note that $F_{\rm stray}$ is slightly larger than
$F_{\rm surf}$; this reflects the fact that the stray field
inside the film is screened by $\lambda' < \infty$, while
outside the film there is no screening ($\lambda = \infty$).
Both $F_{\rm stray}$ and $F_{\rm surf}$ vanish  at $b\to 0$
and at $b \to 1$ and have a maximum in between. At $b\to 0$
one has $F_{\rm stray} \approx F_{\rm surf} \propto b $
since each vortex contributes separately. At $b\to 1$ one has
$F_{\rm stray} \approx F_{\rm surf} \propto (1-b)^2 /\kappa$
(in units $\mu_0 H_c^2 \lambda$) since the amplitude of the
periodic $B_z(x,y, d/2)$ decreases as
$b_{{\bf K}_\perp}^s \propto (1-b)/\kappa$ and the depth of
the stray field is
$K_{10} \approx x_1/2\pi \propto \xi = \lambda/\kappa$.
Therefore, when plotted in units $\mu_0 H_c^2 \xi$, all
curves $F_{\rm stray}$ and all $F_{\rm surf}$ practically
coincide for all $\kappa >1$ and $b > 0.4$, see Fig.~8.

   Note that the total surface energy
$F_{\rm stray} +F_{\rm surf}$ is very small, never
exceeding the value $8 \cdot 10^{-4} \mu_0 H_c^2 \xi$
reached at $\kappa \gg 1$ and $b\approx 0.2/\sqrt{\kappa}$.
 \begin{figure}  
\epsfxsize= .98\hsize  \vskip 1.0\baselineskip \centerline{
\epsffile{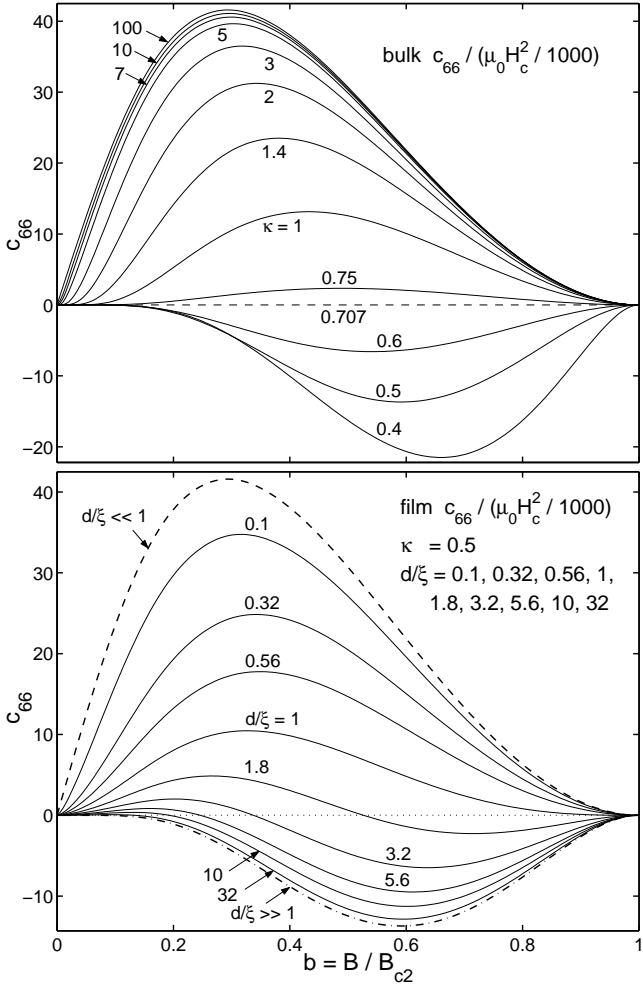}}  \vspace{.1cm}
\caption{\label{fig9}
Top: The shear modulus $c_{66}$ of the bulk ($d\to \infty$)
triangular vortex lattice as function of the reduced
induction $b=\bar B / B_{c2}$
for GL parameters $\kappa = 0.4$, 0.5, 0.6, .707, 0.75, 1,
1.4, 2, 3, 5, 7, 10, 100, in units $\mu_0 H_c^2 / 1000$.
For $\kappa < 2^{-1/2} = 0.707$ one has formally $c_{66} < 0$,
though vortices and a vortex lattice are unstable in such
type-I superconductors.
Bottom: The shear modulus $c_{66}$ of the triangular vortex
lattice in films with thicknesses $d/\xi = 0.1$, 0.32, 0.56,
1, 1.8, 3.2, 5.6, 10, and 32, plotted versus $b$ for
$\kappa = 0.5$. This $c_{66}$ is positive, i.e. the triangular
vortex lattice is stable, for sufficiently thin films or
small inductions. For $d \gg \xi$ the bulk $c_{66}$ at the
same $\kappa =0.5$ is reached (dash-dotted line), and for
$d \ll \xi$ the bulk $c_{66}$ in the limit $\kappa \gg 1$
(dashed line).
 } \end{figure}   
 \begin{figure}  
\epsfxsize= .98\hsize  \vskip 1.0\baselineskip \centerline{
\epsffile{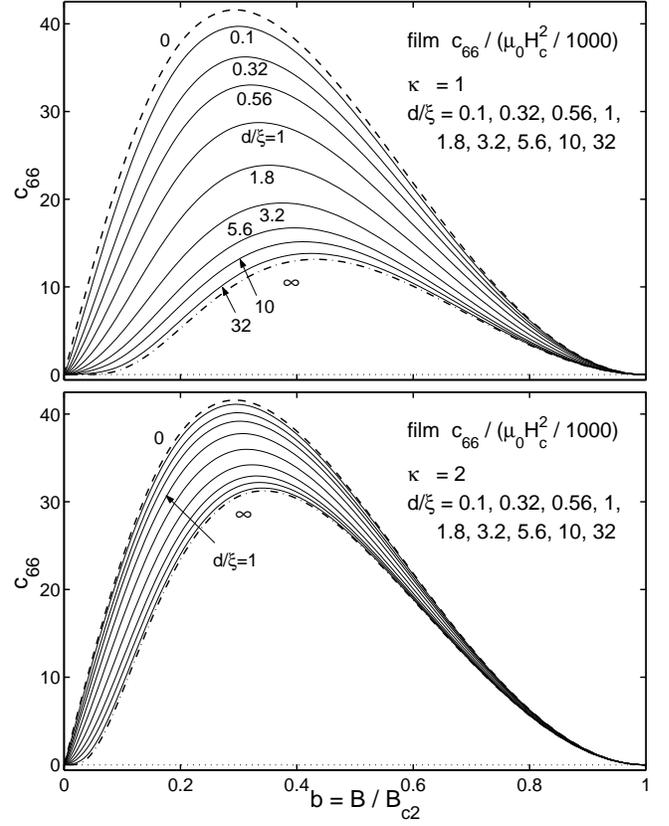}}  \vspace{.1cm}
\caption{\label{fig10}
The shear modulus $c_{66}$ of the triangular vortex
lattice in films of various thicknesses like in Fig.~9 bottom,
but for $\kappa = 1$ (top) and $\kappa = 2$ (bottom),
in units $\mu_0 H_c^2 / 1000$.
 } \end{figure}   

 \begin{figure}  
\epsfxsize= .98\hsize  \vskip 1.0\baselineskip \centerline{
\epsffile{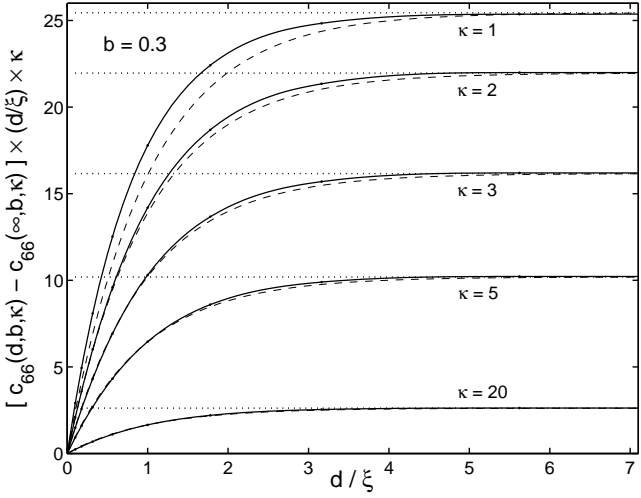}}  \vspace{.1cm}
\caption{\label{fig11}
Dependence of the shear modulus $c_{66}$ of the triangular
lattice on the film thickness $d$. Plotted is the additional
rigidity caused by the film surfaces in form of
$f(d,b,\kappa) =[\,c_{66}(d,b,\kappa) -c_{66}(\infty,b,\kappa)\,]
 \cdot (d/\xi) \cdot \kappa$  versus $d$ at $b=0.3$ for various
$\kappa = 1$ $\dots$  20 (solid lines, $c_{66}$ in units
$\mu_0 H_c^2 / 1000$). The dashed lines show the fit,
Eq.~(32), with $l=\xi$.
 } \end{figure}   

 \begin{figure}  
\epsfxsize= .98\hsize  \vskip 1.0\baselineskip \centerline{
\epsffile{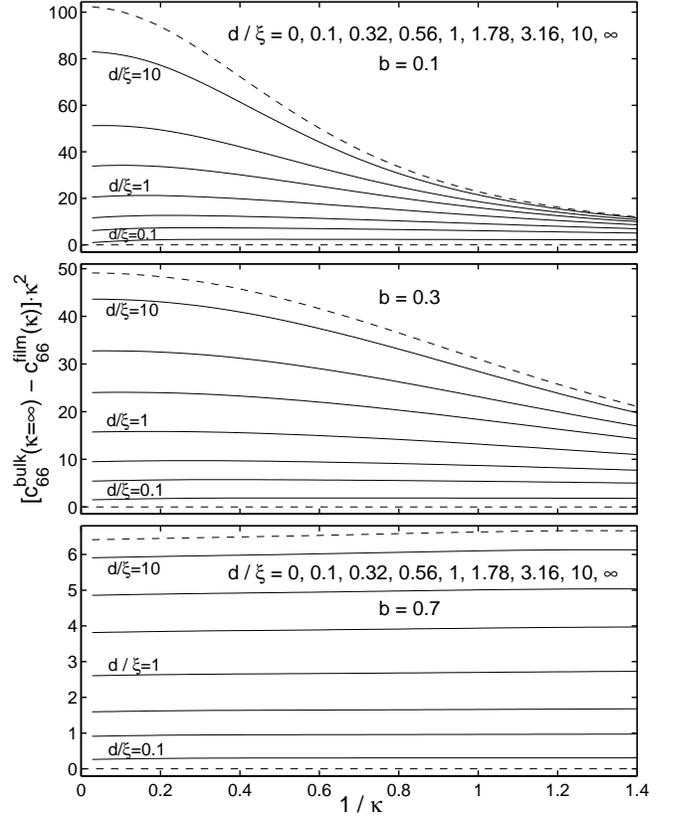}}  \vspace{.1cm}
\caption{\label{fig12}
Dependence of the shear modulus $c_{66}$ of the triangular
lattice on the GL parameter $\kappa$. Plotted is the function
 $[\,c_{66}(\infty,b,\infty) -c_{66}(d,b,\kappa)\,]\cdot\kappa^2$
versus $1/\kappa$ at $b=0.1$ (top), 0.3 (middle), and 0.7 (bottom),
for various film thicknesses $d/\xi$
($c_{66}$ in units $\mu_0 H_c^2 / 1000$). The dashed lines show
the limits $d=0$ ($x$-axis) and $d=\infty$ (upper line) coinciding
with  $[\,c_{66}^{\rm bulk}(\kappa=\infty) -c_{66}^{\rm bulk}
 (\kappa) \,]\cdot\kappa^2$. It is clearly seen that the
differences of any two $c_{66}$ values vanish as $\kappa^{-2}$
when $\kappa \to \infty$. This asymptotic law is good
even for $\kappa \ge 2$, and it practically applies to
all $\kappa \ge 0.71$ at large inductions $b \ge 0.7$.
 } \end{figure}   

 \begin{figure}  
\epsfxsize= .98\hsize  \vskip 1.0\baselineskip \centerline{
\epsffile{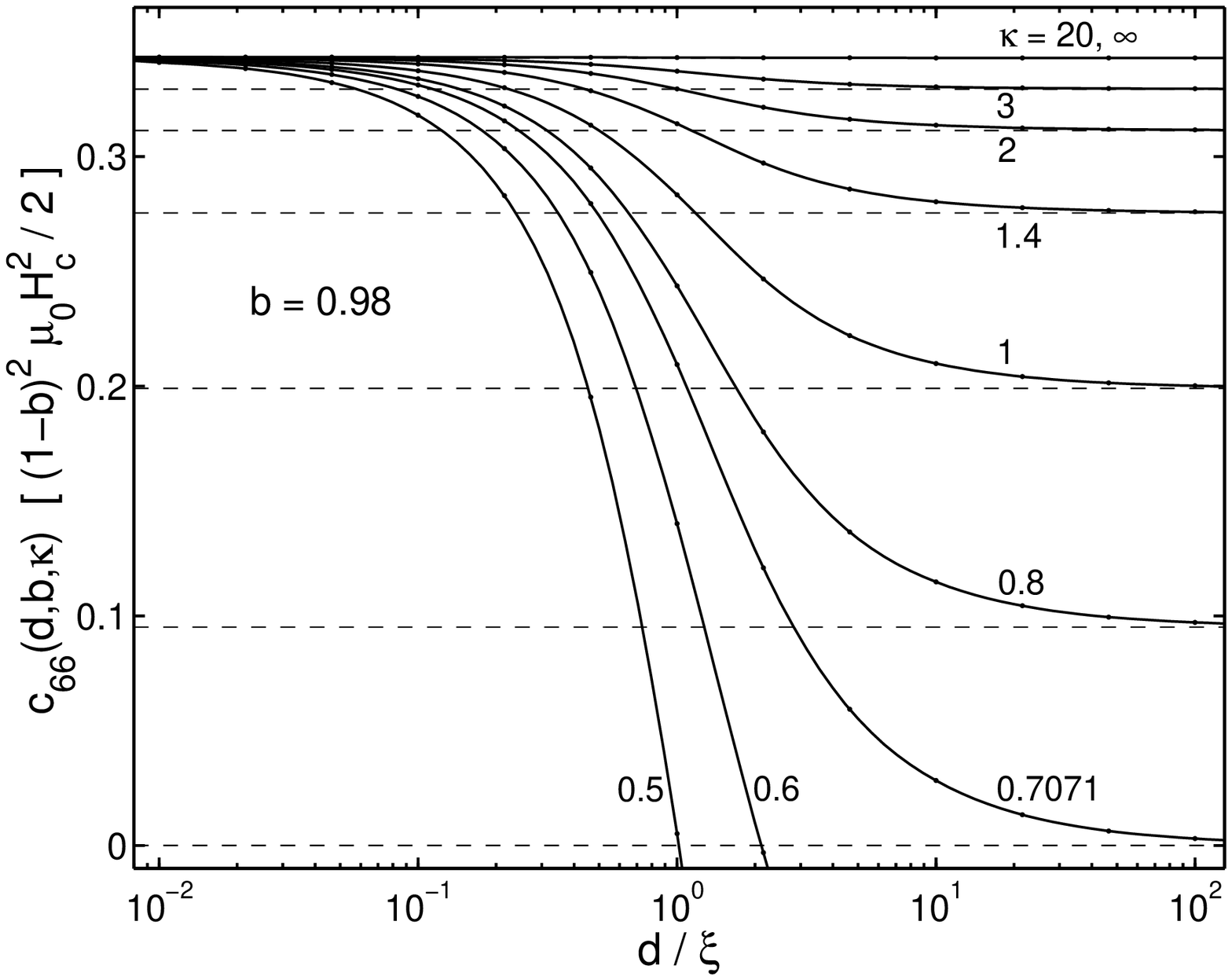}}  \vspace{.1cm}
\caption{\label{fig13}
The shear modulus $c_{66} \propto (1-b)^2$ of the vortex lattice
in films at high inductions $1-b \ll 1$, plotted versus the
film thickness $d$. For comparison with Fig.~1 of
Ref.~\onlinecite{14}, $c_{66}$ is in the same units
$(1-b)^2 \mu_0 H_c^2/2$. The horizontal dashed lines denote the
bulk $c_{66}$. This plot, computed for $b=0.98$, applies
approximately too all $b\ge 0.6$ and agrees with the analytic
expression plotted in Fig.~1 of Ref.~\onlinecite{14}.
 } \end{figure}   

\subsection{Shear modulus of vortex lattice}  

   Finally, the elastic  shear modulus $c_{66}$
of the vortex lattice will be discussed. As shown in
Ref.~\onlinecite{11}, the shear modulus of the triangular
vortex lattice can be expressed with high accuracy by
the difference of the free energies of the rectangular lattice,
$F_{\rm rect}$ (with $x_2=0$ and $y_2 = \sqrt3 x_1 / 2$), and
the  triangular lattice, $F_{\rm tr}$
(with $x_2=x_1/2$ and same $y_2 = \sqrt3 x_1 / 2$),
 \begin{equation}  
    c_{66} = (3\pi^2/2)(F_{\rm rect} -F_{\rm tr}) \,.
  \end{equation}
This is so since the free energy for constant unit cell
height $y_2$ varies practically sinusoidally with $x_2$:
$F(x_2) \approx F_{\rm tr} +[\,1 + \cos(2\pi x_2/x_1)\,]
(F_{\rm rect} -F_{\rm tr})/2$,   thus the definition
$c_{66} = \partial^2 F / \partial \alpha^2$ at small
shear angle $\alpha = \arctan[(x_2 -0.5)/y_2]$ yields
Eq.~(30).\cite{25} Expressed in units $\mu_0 H_c^2$,
the shear modulus depends
on three variables: $c_{66} = c_{66}(b, \kappa, d)$. There
are thus several ways to present the numerical data for
$c_{66}$, each yielding different physical insight.

   One result is that in the limit of small film
thickness $d \ll \xi$ the shear modulus for a film
with any $\kappa$ tends to the bulk shear modulus at
$\kappa \to \infty$, as already obtained by Conen and
Schmid. \cite{14} This finding may be understood from
the fact that in thin films the vortices
are Pearl vortices that have a long interaction range
$2\lambda^2 / d$ exceeding the London penetration depth
$\lambda$.\cite{4,5}  This argument yields the correct
limit $\kappa_{\rm eff} \to \infty$ for
$d / \lambda \to 0$, but for $\kappa < \infty$ the
$c_{66}$ of films does not quantitatively coincide with
the bulk $c_{66}$ for an effective
 $\kappa_{\rm eff} = 2\lambda^2 / d \xi
 = (2\lambda/d) \kappa \gg \kappa$, since $c_{66}$
is determined not only by the range but by the full
shape of the interaction potential between vortices,
which differs for Abrikosov\cite{1} and Pearl\cite{4}
vortices. If this potential is
$V(r)$ with $r = (x^2 + y^2)^{1/2}$
and the vortex density is $n_v = \bar B/\Phi_0$, one has
for a 2D triangular lattice with positions
${\bf R}_{mn}$ defined below Eq.~(3): \cite{26,27}
  \begin{equation}  
  c_{66} = {n_v \over 16}\sum_{m,n}
  [\, R_{mn}^2 V''(R_{mn}) + 3R_{mn} V'(R_{mn})\,] \,,
  \end{equation}
see also Eqs.~(9) and (11) of Ref.~\onlinecite{28}.

  Figures 9 and 10 show $c_{66}$ as a function of the
reduced induction $b$ for different film thicknesses $d$
expressed in units of the GL coherence length $\xi$,
$d/\xi = 10^{(-4, \dots, 6)/4} = 0.1$, $\dots$, 32,
 and for GL parameters $\kappa = 0.5$, 1, and 2. For
large $\kappa \ge 5$, the curves $c_{66}(b)$ for various
thicknesses are very close to each other and to the bulk
$c_{66}$.
In general, the curves for finite film thickness all fall
between the two limiting cases $d \to \infty$ coinciding with
the bulk $c_{66}(\kappa)$, and $d \to 0$ coinciding with the
bulk $c_{66}(\kappa = \infty)$. This interval is very
small for large $\kappa$ and not too small $b$ since
$c_{66}(\kappa=\infty) -c_{66}(\kappa) \propto \kappa^{-2}$.
This means that for large $\kappa \ge 5$ the shear modulus
is nearly the same for thin and thick films. 
Note that for the bulk and $\kappa \ge 5$ one has
$c_{66} \approx B\Phi_0 /(16\pi \mu_0 \lambda^2) \propto b$
for $1/(2\kappa^2) \le b \le 0.15$ (see Fig.~9 top); this
applies also to films. For $\kappa \le 5$ we confirm the
finding of Ref.~\cite{14} that $c_{66} \propto b^{3/2}$ for
$b \ll 1$, but this law applies only to intermediate film
thicknesses $ 0.5 \le d/\lambda \le 3$ at $b \le 0.1$.

  An interesting feature can be seen from Fig.~9. The
upper part shows the {\it bulk} $c_{66}(b,\kappa)$ for
values $\kappa = 0.4$ to $\infty$, i.e., also for
$\kappa < 1/\sqrt{2} =0.707$ corresponding to type-I
superconductors, in which the vortex lattice is
energetically unfavorable in the bulk. For
$\kappa < 0.707$ one finds {\it negative} $c_{66} < 0$.
This means the bulk rectangular and square vortex
lattices \cite{25} have lower energy than the triangular
lattice (the Meissner state without vortices has even
lower energy in this case).
However, as can be seen in the lower plot for films
with $\kappa = 0.5$, in sufficiently thin type-I
superconductor films the triangular vortex lattice can
be stable (i.e., $c_{66} >0$) when the induction is
sufficiently small. This behavior was seen also in
Ref.~\onlinecite{14}.

  The dependence of $c_{66}$ on the film thickness $d$
is visualized in Fig.~11 by plotting the difference
  \begin{equation}  
  f(d,b,\kappa) = [\, c_{66}(d,b,\kappa) - c_{66}
  (\infty,b,\kappa) \,] \cdot (d/\xi) \cdot \kappa
  \end{equation}
(an energy per unit area) versus $d$ at various
$\kappa$ values for $b=0.3$ (near the maximum of
$c_{66}$). One can see that this function saturates
when the film thickness exceeds a few coherence lengths
$\xi$. For all values of $\kappa$ and $b > 1/\kappa^2$
one can fit these curves closely by
  \begin{equation}  
  f(d) \approx f(\infty) [\,1 - \exp(-d/l) \,] \,.
  \end{equation}
In Fig.~11 (at $b=0.3$) the length $l$ of the best fit
accidentally coincides with $\xi$, but in general $l$
depends on $b$ and is proportional to the vortex
spacing $x_1$:  $l \approx 0.195 x_1 = (\sqrt6 / 4\pi)
 x_1 = \sqrt2 / K_{10}$, thus $l/\xi \approx 3^{1/4}
 (2\pi)^{-1/2}/\sqrt{b} =0.525/ \sqrt{b}$, yielding
 $l = 0.96\xi$ at $b=0.3$. This saturation means that
the additional rigidity of the vortex lattice caused by the
film surfaces and measured by $f(d,b,\kappa)$, becomes
independent of $d$ in films thicker than a few coherence
lengths, and thus one has
$c_{66}(d) - c_{66}(\infty) \propto 1/d$. For thin films
with $d \ll \xi$ one has $f(d,b,\kappa) \propto d$ since
$c_{66}(d) - c_{66}(\infty)$ is a constant independent
of $d$.

   Figure 12 visualizes the $\kappa$ dependence of the
shear modulus of the triangular lattice by plotting
$[\,c_{66}(\infty,b,\infty) -c_{66}(d,b,\kappa)\,]\cdot\kappa^2$
versus $1/\kappa$ at $b=0.1$, 0.3, and 0.7, for film
thicknesses $d/\xi = 0.1$ $\dots$ 10. Also shown are the
limits $d=0$ (the $x$-axis), proving that
$c_{66}(d \to 0,b,\kappa) =c_{66}(d\to\infty,b,\kappa\to\infty)$
for any $\kappa$,
and $d=\infty$ (upper dashed line) that coincides with
$[\,c_{66}^{\rm bulk}(\kappa=\infty) -c_{66}^{\rm bulk}
 (\kappa) \,]\cdot\kappa^2$. These plots prove that the
differences of any two $c_{66}$ values vanish as $\kappa^{-2}$
when $\kappa \to \infty$. One can see that this asymptotic
law is a good approximation even for not so large $\kappa\ge 2$,
and at large $b \ge 0.7$ it practically applies to all
$\kappa \ge 0.71$, and even for type-I superconductors with
not too small $\kappa < 0.71$.

   In Fig.~13 the numerical $c_{66}(d,b,\kappa)$ is compared
with the analytical result of Conen and Schmid, Fig.~1 of
Ref.~\onlinecite{14}, valid at large inductions $1-b \ll 1$.
Their result was obtained from an elegant expression
derived by Lasher \cite{18} for the free energy of films
with vortex lattice of arbitrary symmetry at $b \to 1$.
Lasher \cite{18} implicitly used the fact that the magnetic
stray field inside the film is not screened in this limit
of $b \to 1$.
 \begin{figure}  
\epsfxsize= .98\hsize  \vskip 1.0\baselineskip \centerline{
\epsffile{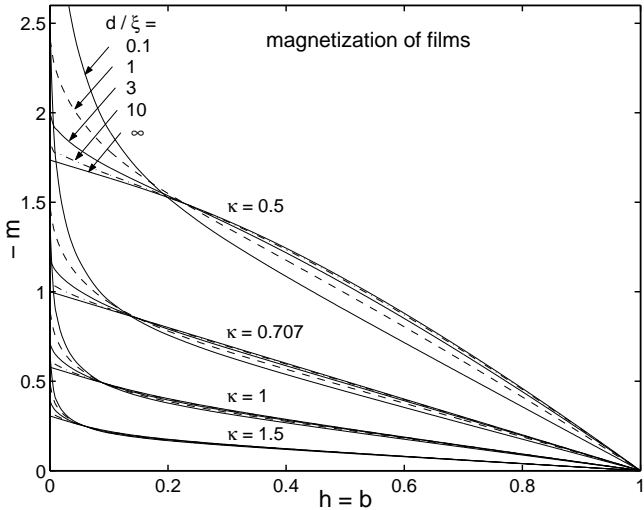}}  \vspace{.1cm}
\caption{\label{fig13}
The magnetization of infinite films with thickness
$d/\xi = 0.1$, 1 ,3, 10, $\infty$, containing a triangular
vortex lattice with one flux quantum per vortex. Plotted is
$-m =-M / H_{c2}$ versus $h = H/H_{c2} = b = B/B_{c2}$ for
$\kappa = 0.5$, 0.707, 1, 1.5.
 } \end{figure}   

\subsection{Magnetization of infinite films}  

   For infinitely extended films the average magnetic induction
$\bar B$ equals the applied field, $\bar B = \mu_0 H_a$,
and the magnetization $M$ is defined as the magnetic moment per
unit volume of the film. The demagnetization factor of infinite
films is $N=1$, and thus the effective lower critical field at
which the first vortices penetrate is $H_{c1}' = (1-N)H_{c1}=0$.
For the magnetization of superconductors with general
demagnetizing factor $0 \le N \le 1$ see, e.g.,
Refs.~\onlinecite{19,29}. Noting that the total free energy
per volume $F_{\rm tot}$, Eq.~(9), equals the work done by
the applied field, which may be written as $- \int M\, dB$, one
obtains the relation $M = -\partial F_{\rm tot} / \partial B$.
Figure 14 shows magnetization curves for films of various
thicknesses $d/\xi$ = 0.1, 1, 3, 10, and $\infty$
for GL parameters $\kappa = 0.5$, $1/\sqrt 2$, 1, and 1.5.

   For thick films ($d \gg \xi, \lambda$) at
 $\kappa = 1/\sqrt 2$ one has
 $F_{\rm tot} = -{1 \over 2}(1-b)^2$ and thus $-m=1-b=1-h$;
for larger $\kappa > 0.707$ the thick film $-m(h)$ is concave
(has positive curvature);
and for smaller $\kappa < 0.707$ (type I superconductors)
$-m(h)$ is convex (has negative curvature) and the vortex
lattice is not stable. However, for sufficiently thin films,
even when $\kappa < 0.707$ the curvature of $-m(h)$ can be
positive and even may change sign at a certain value of $h=b$.
Note that the plotted curves $-m(h)$ for various $d/\xi$ cut
each other approximately at $h=b \approx 1/\kappa$. For thick
films the initial slope is $-m'(h)|_{h=0} = -1$ for all
$\kappa$, and $-m(0) = h_{c1} = H_{c1} / H_{c2}$. This is so
since when surface contributions may be disregarded, one has
for superconducting ellipsoids $-M = H_{c1}$ at $H=H'_{c1}$
where $B=0$. The enhancement of $-m(h)$ at small $h < \kappa$
for thin films in Fig.~14, originates from the energy of the
magnetic stray field, which enhances the self energy of a
vortex beyond its bulk value $d \Phi_0 H_{c1}$.
 More details about this will be published elsewhere.

\section{Summary}  

   It is shown how the Ginzburg-Landau equations can be solved
for a periodic lattice of Abrikosov vortices in superconducting
films in a perpendicular magnetic field. As illustration how
well this iteration method works, some results are presented.
The widening of the magnetic field lines as they exit the
film surface is correctly obtained, Figs.~1, 2, but this
leads only to a very small correction of the order
parameter near the surface, Fig.~1. The variance of the
transverse component of the magnetic induction is sharply
peaked at the surface and vanishes deep inside and far outside
the film, Figs.~4, 5. The variance of the periodic order
parameter (Abrikosov parameter $\beta$) varies very little
across the film thickness, by at most a factor 1.0033, Fig.~6.
   The surface energy saturates for large film
thickness $d$ and vanishes linearly at small $d$, Fig.~7.
For not too thin films the surface energy originates mainly
from the magnetic stray field, which comes in approximately
equal parts from outside and inside the film, in particular
for large $\kappa$ or large $b$, Fig.~8. For very thin films
the stray field energy may be disregarded and the very small
surface energy comes mainly from the small modification
of the order parameter, Fig.~7.
The shear modulus $c_{66}(d,b,\kappa)$ of the triangular
vortex lattice in thin films approaches  the
$c_{66}(\infty,b,\infty)$ of thick films (bulk limit) at
$\kappa \to \infty$, Fig.~10, the difference being
proportional to $\kappa^{-2}$, Fig.~12. While the bulk
$c_{66}(\infty,b,\kappa)$ is negative in type-I superconductors
($\kappa < 0.707$), the $c_{66}$ of sufficiently thin films can
be positive and may change sign at some value of $b$, Fig.~9.
The magnetization curves of films with $\kappa < 0.707$
may have positive or negative curvature, depending on the film
thickness, Fig.~14. More results will be published elsewhere.
Extensions of this method to vortices with several flux quanta
and to the periodic lattice of curved vortices in superconducting
films in a tilted magnetic field are underway.
\vspace{5 mm}

\appendix
\section{Presentation of field lines}  

  A practical question is how to plot the magnetic field
lines of this 3D problem such that they have the correct
orientation of ${\bf B}(x,y,z)$ and their 1D density
(reciprocal distance) in the plotted plane is proportional
to the magnitude $|{\bf B}(x,y,z)|$. A simple
consideration shows that this is possible only for 2D
planar problems, when the field lines coincide with the
contour lines of the vector potential, e.g., $A_y(x,0,z)$.
But for 3D magnetic fields, and even for cylindrically
symmetric fields, such 2D plots of the field lines are
not possible since the magnitude $|{\bf B}(x,y,z)|$
here is proportional to the 2D density of the 3D
field lines, but not to the 1D density of the plotted
2D field lines. For our 3D problem of a thick film
with a 2D periodic vortex lattice we have two
possibilities to plot field lines that approximately
have the above mentioned properties.

  {\em First method:\/} One may use numerical programs
that plot the field lines (stream lines) of the 2D planar
field ${\bf B}(x,0,z)= (B_x, B_z)$ [or any other
planar cross section of ${\bf B}(x,y,z)$] starting from
equidistant points ($x=x_i, y=0, z=-z_0$) far away from
the film surface so that
${\bf B}(x,y,z) \approx \bar B {\bf \hat z}=$ const.
Such field lines have the correct slope, but their
density is only approximately proportional to
$|{\bf B}(x,y,z)|$.

   {\em Second method:\/} In this paper the 2D plots
of the magnetic field lines show the contour lines of
the function
  \begin{eqnarray}  
\varphi(x,z) = \int_0^x\!\! B_z(x,0,z)\,dx\bigg/ \!
               \int_0^{x_1}\!\!\!\! B_z(x,0,z)\,dx \,,
  \end{eqnarray}  \\[-2mm]
which ranges from $\varphi(0, z) =0$ at $x=0$ (vortex
center) to $\varphi(x_1/2, z) =1$ (middle plane
between two vortices) and has a periodic derivative.
Such field lines have a 1D density along $x$ proportional
to $|B_z(x,0,z)|$, and a density perpendicular to
these lines which is close to
$|{\bf B}(x,y,z)|$, since their orientation is
close to the orientation of ${\bf B}(x,y,z)$.
Figure~2 shows that these two types of field lines
are very similar. In particular, the contour lines
of $\varphi(x,z)$, Eq.~(1), have slopes that are
close to the correct slope.



\begin{thebibliography}{9}
\bibitem{25} In Eq.~(5) of Ref.~\onlinecite{12} erroneously
             $F_{\rm sq}$ (square lattice) was written in place of
             $F_{\rm rect}$. This form of Eq.~(30) becomes
             numerically correct if the factor $(3\pi^2 /2) = 14.8$
             is replaced by 23.1, i.e., $F_{\rm sq}$ lies
             approximately in the middle between
             $F_{\rm tr}$ and $F_{\rm rect}$.
\end{thebibliography}
\end{document}